# On the singular behaviour of structure functions at low x

by

H. Navelet, R. Peschanski, S. Wallon

Service de Physique Théorique, CEA-Saclay

F-91191 Gif-sur-Yvette Cedex, FRANCE

**ABSTRACT**

We discuss the phenomenological extraction of the leading $j$-plane singularity from singlet structure functions $F_s$ measured at small $x$. Using a saddle-point method we show that $\mathrm{dln}F_s/\mathrm{dln}\frac{1}{x}$ is a suitable observable for this purpose in the region $x \le 10^{-2}$. As an application, we confront and distinguish in a model-independent way structure function parametrizations coming from two different QCD evolution equations, namely the Lipatov (or BFKL) equation and the Gribov-Lipatov-Altarelli-Parisi (or GLAP). Recent results at HERA are discussed in this framework.

Recent results from HERA experiments ($H_1$[1] and Zeus[2]) yield information on the singlet proton structure function $F_s$ for a new range in the Bjorken variables, namely $10^{-4} \lesssim x \lesssim 10^{-2}$, $8 \lesssim Q^2 \lesssim 60$ GeV$^2$. These experimental results, although of low statistics, have already provided promising theoretical information[3].

The rise of $F_s$ as $x$ becomes very small, has reinforced predictions of parametrizations based on perturbative QCD evolution equations, of a large increase in space-like gluon radiation. The QCD calculation in the leading $\left(\alpha_s \mathrm{ln}\frac{1}{x}\right)^n$ approximation[4] appears to be relevant for this phenomenon, even though non-perturbative effects[5] or gluon-gluon interactions[6] may modify its quantitative predictions.

In this framework, the structure function is dominated by a singularity in the $j$-plane of its Mellin transform, namely

$$F_s(x) \sim x^{-a} \left(\ln\frac{1}{x}\right)^b$$

$$\int_0^1 \mathrm{d}x\ x^{j-1} F_s(x) \sim \tilde{F}_s(j) \sim \frac{1}{(j-a)^{b+1}} \tag{1}$$

with $a > 1$ being the end-point singularity, and $b$ characterizing the type of singularity (square root $b = -\frac{1}{2}$, pole $b = 0$, etc...). An exact calculation of the Lipatov (BFKL) singularity at a *fixed* external momentum scale yields $b = -\frac{1}{2}$ but this coefficient may be substantially different for structure functions[5].

The Lipatov singularity is given by $a = 1 + 4\ln 2 \frac{N_c}{\pi}\bar{\alpha}_s$ and may reach 1.5. However, even though this perturbative approach looks promising, some ambiguities have still to be resolved. For instance, non-perturbative contributions are present because of the large diffusion in the gluon transverse momentum scales contributing to the structure function[5]. More intringuingly a prediction (GRV parametrization[7]) based on the GLAP evolution equation[8] has also been successful at HERA. Its success is surprising because it is based on extrapolating from a very small value of $Q^2 = \mu^2$ by summing only the leading $\left[\alpha_s \ln Q^2\right]^n$ contributions, and thus on a different picture of the $j$-plane singularity. Is its success a mere coincidence or is it due to a subtlety within the theory? This is the question we wish to settle.

Our aim in this letter is to answer this question by proposing a model-independent study of the $j-$plane effective singularity of the structure functions. Our results are as follows:

i) When using a saddle-point approximation valid at small $x$ ($x \lesssim 10^{-2}$), the Mellin transform of eq.(1) reduces to a local correspondence. $x \longrightarrow \bar{j}(x)$ between $F_s$ and $\tilde{F}_s(\bar{j})$.

ii) Within the same approximation, the function $\mathrm{d}\ln F_s(x)/\mathrm{d}\ln\frac{1}{x}$ depends only on the rightmost $j-$plane singularity of $\tilde{F}_s$. This result is in agreement with a recent empirical observation[9].

iii) Applying this result to the BFKL or GLAP $j-$plane singularity, we show that while different analytically, they lead to an almost identical $\mathrm{dln}F_s(x)/\mathrm{dln}\frac{1}{x}$ in the small $x$ range of the HERA experiments, at fixed $Q^2$.

iv) The only way to distinguish the two types of singularities may be the $Q^2$ dependence of $\mathrm{dln}F_s(x)/\mathrm{dln}\frac{1}{x}$.

Let us prove assertion i). The saddle-point approximation relating a structure function at small $x$ and its Mellin transform near its rightmost singularity is correct for the Lipatov singularity[11]. We will use this method in a general way. Let us consider the inverse Mellin transform of relation (1)

$$F_s(x) = \frac{1}{2i\pi}\int_{\mathbb{C}} \mathrm{e}^{\,j\ln 1/x + f}\,\mathrm{d}j \tag{2}$$

where $f \equiv \ln\tilde{F}_s(j)$ and $\mathbb{C}$ is a contour in the complex $j$-plane with Re $j = j_0$ at the right of all singularities of $\tilde{F}_s(j)$.

For $\ln 1/x$ large enough, we get using the saddle-point approximation

$$\ln F_s(x) \simeq \bar{j}\,\ln\frac{1}{x} + f - \frac{1}{2}\ln(2\pi f''), \tag{3}$$

where the saddle point $\bar{j}(x)$ is defined by the condition

$$0 = \ln\frac{1}{x} + f'(\bar{j}) \tag{4}$$

and the primes mean the derivative w.r.t. $j$ at $j = \bar{j}$, i.e. $f' \equiv \left.\frac{\mathrm{d}f}{\mathrm{d}j}\right|_{\bar{j}}$. Equation (4) gives a local relation $x \longrightarrow \bar{j}(x)$. Moreover, $f'$ is given by the dominant $j$-plane singularity of $\tilde{F}_s(j)$. As an example the singularity given in eq.(1) yields $\bar{j} = a + (b+1)/\ln\frac{1}{x}$. The value of $\left[\ln\frac{1}{x}\right]^{-1}$ is a measure of the distance to the singularity at $j = a$.

Next we prove assertion ii) about $\mathrm{dln}F_s/\mathrm{dln}\frac{1}{x}$. Using the saddle-point formula (3), we get by differentiating,

$$\frac{\mathrm{d}\ \ln\tilde{F}_s}{\mathrm{d}\ \ln\frac{1}{x}} = \bar{j} + \frac{\mathrm{d}\bar{j}}{\mathrm{d}\ \ln\frac{1}{x}}\left\{\ln\frac{1}{x} + f' - \frac{1}{2}\frac{f'''}{f''}\right\}.$$

Note that $\frac{\mathrm{d}\ \bar{j}}{\mathrm{d}\ \ln\frac{1}{x}} = [f'']^{-1}$. Finally, using (4), the derivative reads

$$\frac{\mathrm{d}\ \ln\tilde{F}}{\mathrm{d}\ \ln\frac{1}{x}} = \bar{j} - \frac{1}{2}\left[f''^{-1}\right]', \tag{5}$$

(this simple expression stems from the fact that the structure function is a Legendre transform for the conjugate variables $j$ and $\ln\frac{1}{x}$, see eq.(2)). Now we note that eq.(5) is dominated by the rightmost $j$-plane singularity of $\tilde{F}_s$. Indeed, non-singular (or less singular) contributions appears only in the saddle-point correction term, namely $\left[f''^{-1}\right]'$, and they are completely negligible in the small $x$ domain.

Indeed, a similar observation has been made recently about a background-subtracted structure function[9]. Our derivation, using the saddle-point method, provides an explanation for this empirical observation. The background has a singularity near $j = 1$, which has to be subtracted out in order to extract the main BFKL contribution as done in ref. [9].

Now let us consider the well-known $j-$plane singularity generated by the GLAP evolution equations[8]. Define our notation by

$$\begin{aligned}\tilde{F}_s(j) &= \left(\frac{\nu_F - \nu_-}{\nu_+ - \nu_-}\exp\ \nu_+\xi + \frac{\nu_+ - \nu_F}{\nu_+ - \nu_-}\exp\ \nu_-\xi\right)(q+\bar{q})(j) \\ &+ \frac{2N_f\phi_G^F}{\nu_+ - \nu_F}\left(\exp\ \nu_+\xi - \exp\ \nu_-\xi\right)g(j) \\ &\equiv \frac{\nu_F - \nu_-}{\nu_+ - \nu_-}\exp\ \nu_+\xi\ *\ (\text{a regular function at } j > 0),\end{aligned} \tag{6}$$

where $\xi$ is the well known evolution parameter of the GLAP equation and $(q + \bar{q})$ (resp. $g$) are the input quark singlet (resp. gluon) structure functions at $\xi = 0$. Note that here we want to analyze the singularity structure of the GLAP evolution equation. Thus, we will apply our formalism to the case of input functions $q, \bar{q}, g$ which are *regular* for $j > 0$.

The GLAP kernels read

$$\nu_F = -\frac{4}{3}\left\{4\psi(j+1) + 4\gamma_E - 3 - \frac{2}{j(j+1)}\right\}$$

$$\nu_G = -12\left\{\psi(j+1)+\gamma_E\right\} + 11 - \frac{2}{3}N_f + \frac{24\left(j^2+j+1\right)}{j\left(j^2-1\right)(j+2)}$$
$$\phi_G^F = \frac{j^2+j+2}{j(j+1)(j+2)} \qquad \phi_F^G = \frac{8}{3}\frac{j^2+j+2}{j\left(j^2-1\right)}$$
$$\nu_\pm = \frac{1}{2}\left\{\nu_F + \nu_G \pm \sqrt{\left(\nu_F-\nu_G\right)^2 + 8N_f\phi_F^G\phi_G^F}\right\}, \tag{7}$$

where $\psi$ is the logarithmic derivative of the $\Gamma$ function and $\gamma_E$ the Euler constant.

At this stage, some comments are in order. The full leading singularity at $j = 1$ in the GLAP equations is taken into account since all singular contributions at $j = 1$ (the essential singularity contained in $e^{\nu_+\xi}$ and its polynomial prefactor) are present. This is due to the fact that $\nu_F, \nu_-, \phi_G^F$ are non-singular at $j = 1$ whereas $\nu_+ = \frac{12}{j-1}$+regular function at $j > 0$.

On the other hand, the whole singularity structure of the GLAP equation depends only on the parameter $\xi$

$$\xi = \frac{1}{11-\frac{2}{3}N_f}\ln\left\{\frac{\ln\ Q^2/\Lambda^2}{\ln\ \mu^2/\Lambda^2}\right\}, \tag{8}$$

where $Q$ is the scale of the deep-inelastic reaction, $\mu$, the initial evolution scale, $\Lambda$, the QCD one-loop scale and $N_f$ the number of flavors. Note that a phenomenological application[7] may require evolution beneath the perturbative domain i.e. ($\mu < Q \leq Q_0$), where $Q_0^2 \sim$ GeV$^2$. In our study, we shall use $\xi$ to parametrize the GLAP singularity including the non-perturbative evolution range.

Now let us demonstrate assertion iii), i.e. the equivalence between the GLAP and BFKL singularities in the small $x$ range at fixed $Q^2$. In Fig.1, the predictions of $d\ln F/d\ln\frac{1}{x}$ are given for three typical values of $\xi$ and compared with the 2-parameter fit using the BFKL expression (1). The $a$ and $b$ parameters of eq.(1) are displayed as a function of $\xi$ in Fig.2. Surprisingly enough, although the singularities are different analytically, both give similar results in the region of interest for HERA, $10^{-4} \lesssim x \lesssim 10^{-2}$. We have performed a variety of checks on this claim, varying the fitting range, and switching on and off the non-singular terms in the GLAP kernels themselves. This results in a small spread in the values of $a$ and $b$, see Fig.2. For $x \geq 10^{-2}$ the regular terms can no longer be neglected, while for $x \leq 10^{-4}$ the difference between the two parametrizations becomes noticeable, though still small.

Now let us come to our final assertion iv) about the $Q^2$-evolution of $\mathrm{dln}F/\mathrm{dln}\frac{1}{x}$. The BFKL singularity is not expected to vary significantly with $Q^2$, even if we take into account the running of the coupling constant[11]. To be more specific, the overall normalization of the singlet structure functions is indeed expected to vary as $\sqrt{Q^2}$ but its shape is expected to remain unchanged[12]. This implies that $\mathrm{dln}F/\mathrm{dln}\frac{1}{x}$ is independent of $Q^2$. In contrast, a distinctive feature of the GLAP singularity is its $Q^2$ evolution. For the quantity $\mathrm{dln}F/\mathrm{dln}\frac{1}{x}$ this evolution is determined by the $\xi$ dependence of $a$ and $b$, see Fig.2. Note that a $Q^2$ dependence compatible with the first-order GLAP equations has also been obtained in phenomenological studies of parton distributions evolution at small $x$, especially when the input functions are singular[13,14].

We will first discuss the phenomenological implications of our remarks. In Fig.3, we exhibit the data with corresponding fits for $Q^2 = 15$ and 30 $(\mathrm{GeV/c})^2$, values which correspond to the best-explored small-$x$ domain. The lack of statistics in the present data excludes the possibility of obtaining a precise estimate of $\mathrm{dln}F_s/\mathrm{dln}\frac{1}{x}$. For this reason, we will proceed as following. We consider the scale $\mu$ in eq. (8), as the only relevant parameter with $\Lambda$ fixed at a reference value of 100 MeV. Then, leaving free the normalization of the structure function at each $Q^2$, we obtain an estimate of $\mu = 290 \pm 60$MeV. This preliminary analysis is an incentive for a more complete study which will be allowed by the larger statistics at HERA in the near future. At the very least, it confirms that either a BFKL or a GLAP singularity may well describe the present data. In agreement with the GRV phenomenological analysis[7], the value of $\mu$ is situated in the non-perturbative range of scales for QCD.

Some theoretical comments are now in order.

**The BFKL singularity**

This singularity is obtained as a resummation of the leading $\left[\alpha_s \mathrm{ln}\frac{1}{x}\right]^n$ perturbative

QCD contributions coresponding to soft gluons. The measurement of proton structure functions in the low-$x$ region accessible at HERA, may provide the first experimental signal of the BFKL singularity predicted long ago from QCD calculations[4]. However it is also known that this singularity contains non-perturbative contributions connected to the low transverse-momentum range of the gluons in the corresponding ladder diagrams. This is often cited as the $k_T-$diffusion problem[5]. A somewhat related problem is the the infrared cut-off dependence[15] of the BFKL singularity, especially when the coupling constant is allowed to run. Let us see what can be learned from the comparison with the GLAP evolution equations.

**The GLAP singularity**

This singularity is obtained as a resummation of the leading $\left[\alpha_s \ln Q^2\right]^n$ perturbative QCD contributions corresponding to the resummation of collinear gluons. More precisely, the use of this singularity at small $x$ is nothing but the double logarithm approximation (DLA). The DLA approximation is not expected to be dominant in the present range of HERA experiments Thus it came as a surprise that the GRV parametrization[7] based on a fit of previous experimental results using the GLAP evolution equations over a large range of scales yields a good prediction of HERA data[1,2]. We now see by our method using $d\ln F/d\ln\frac{1}{x}$ that it is not a mere coincidence but a consequence of the striking similarity between GLAP and BFKL singular behaviour in the experimental small-$x$ range.

We obtained our results for evolution equations including a non-perturbative range, $\mu \leq Q \leq Q_0$. It is worthwhile to compare them to those obtained some time ago[13] in the deep-perturbative regime, $Q \gg Q_0$. Starting with the same input at $Q_0$, the BFKL and GLAP evolution of the structure functions is found to be quite similar. The theoretical origin of this behaviour is the approximate equality of gluon anomalous dimensions when the expansion parameter $\frac{\alpha_s(Q^2)}{j-1}$ is small[13]. In our analysis, the expansion parameter near $j=1$ is $\frac{a-1}{j-1}$, see eq. (1), remaining of order 1 in the domain of phenomenological interest. Our conclusion is that the end-point of the singularity $a$ is compatible with a genuine

BFKL singularity, while the type of singularity is modified. This can be seen from Fig.(2), where the parameter $b$ evolves from a value not far from the predicted one ($b \sim -\frac{1}{2}$), towards positive values at larger $\xi$.

One may object to the validity of the GLAP formalism in a region which includes non-perturbative contributions[16]. But, as we have already mentioned, non-perturbative contributions also have to be taken into account in the BFKL formalism for deep inelastic structure functions[5]. There may well be less difference between the two approaches at small $x$ than expected. Let us discuss this point a little further. If this similarity is more than a coincidence, one obtains a new information on the *non-perturbative* region of the $k_T-$diffusion of the BFKL gluon radiation. This suggests that the dominant $\ln\frac{1}{x}, \ln k_T$ ranges for gluons in the ladder diagrams[5] contributing to $F_s$ (assuming the extension of the BFKL evolution equation to the non-perturbative domain) could be in one-to-one correspondence with those in the GLAP equations with an evolution from a low $\mu^2$ scale. The hadronic final states in Deep-Inelastic Scattering could be a sensitive test of this conjecture.

As a final remark concerning the $Q^2$ dependence, there seems to be a slight theoretical contradiction between the predictions of the BFKL and GLAP evolution equations at small $x$. Indeed, the latter can be proven to be valid if the input structure functions are singular enough[13,14] while the former predicts a different $Q^2$-evolution pattern[12]. We have found a specific $Q^2$ evolution of the BFKL parameters $a(\xi), b(\xi)$ which remains to be studied theoretically and phenomenologically in order to clarify this point.

**Acknowledgements**

We thank J. Baudot warmly for his contribution to the initial stage of this work. Thanks also to M. Besançon, J.-F. Laporte and Ch. Royon for fruitful discussions and A.D. Martin for communicating the paper of ref.[9].

# FIGURE CAPTIONS

Figure 1: **Comparison of BFKL and GLAP singularities for** $\mathrm{dln}F/\mathrm{dln}\frac{1}{x}$

The predictions for $\mathrm{dln}F/\mathrm{dln}\frac{1}{x}$ in the small-$x$ range are displayed for the GLAP singularity (dotted line) and the BFKL singularity (continuous line). The small-$x$ range is limited from above (excluding the dashed lines) by the range of validity of the $j-$plane singularity dominance (see text). The curves are displayed for three typical values $\xi = (.1, .2, .3)$ of the GLAP evolution parameter. The corresponding values of $Q$ span a very large range between 1.7 and $810^5$GeV(!) with the chosen parameters (see formula (8) in the text).

Figure 2: **Correspondence between BFKL and GLAP singularity parameters**

The $a$ and $b$ parameters of the BFKL singularity, eq.(1), are displayed as functions of $\xi$, the GLAP evolution parameter (full black stripes). The width corresponds to systematic variations on the fitting range, see text. The grey region corresponds to a simple power-law fit ($b \equiv 0$).

Figure 3: **Phenomenological comparison**

The data for the singlet structure function at small $x$ from $H_1$ ($Q^2 = 15, 30$ GeV$^2$ from ref.[1]) and Zeus ( $Q^2 = 30$ GeV$^2$ from ref.[2]) are displayed in a $\ln F_s / \ln\frac{1}{x}$ plot. The continuous line corresponds to a fit with $\mu = 290$MeV, see formula (8). (The absolute normalization is fixed independently for each set of data, since we want to parametrize only the derivative $\mathrm{dln}F/\mathrm{dln}\frac{1}{x}$). An estimate of the admissible values give $\mu = 290 \pm 60$MeV, which correspond to $\xi = .135 \pm .015$ ( $Q^2 = 15$ GeV$^2$) and to $\xi = .15 \pm .02$ ( $Q^2 = 30$ GeV$^2$), all other parameters being fixed as in the text.

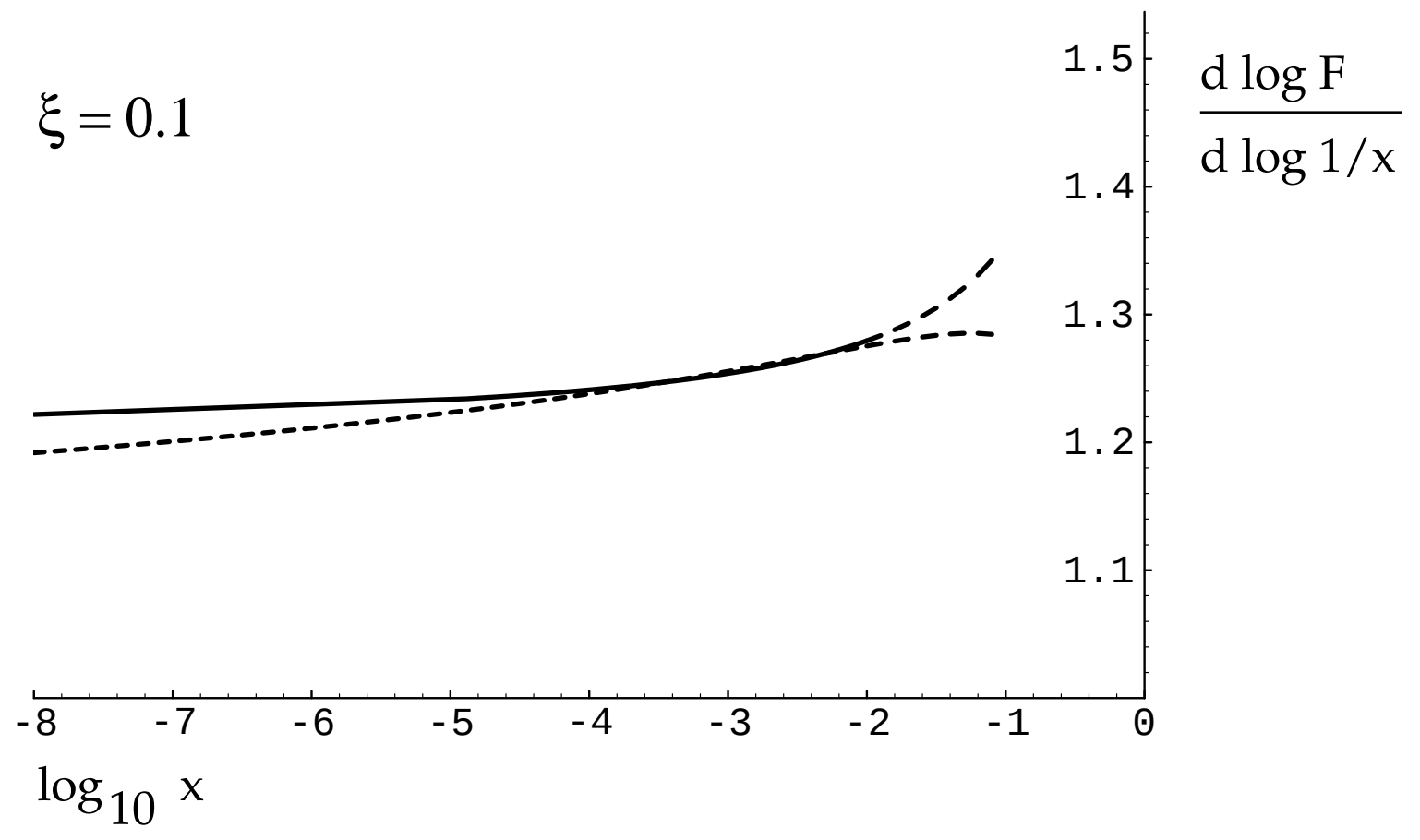
ξ = 0.1
1.5
1.4
1.3
1.2
1.1
d log F / d log 1/x
-8 -7 -6 -5 -4 -3 -2 -1 0
log10 x

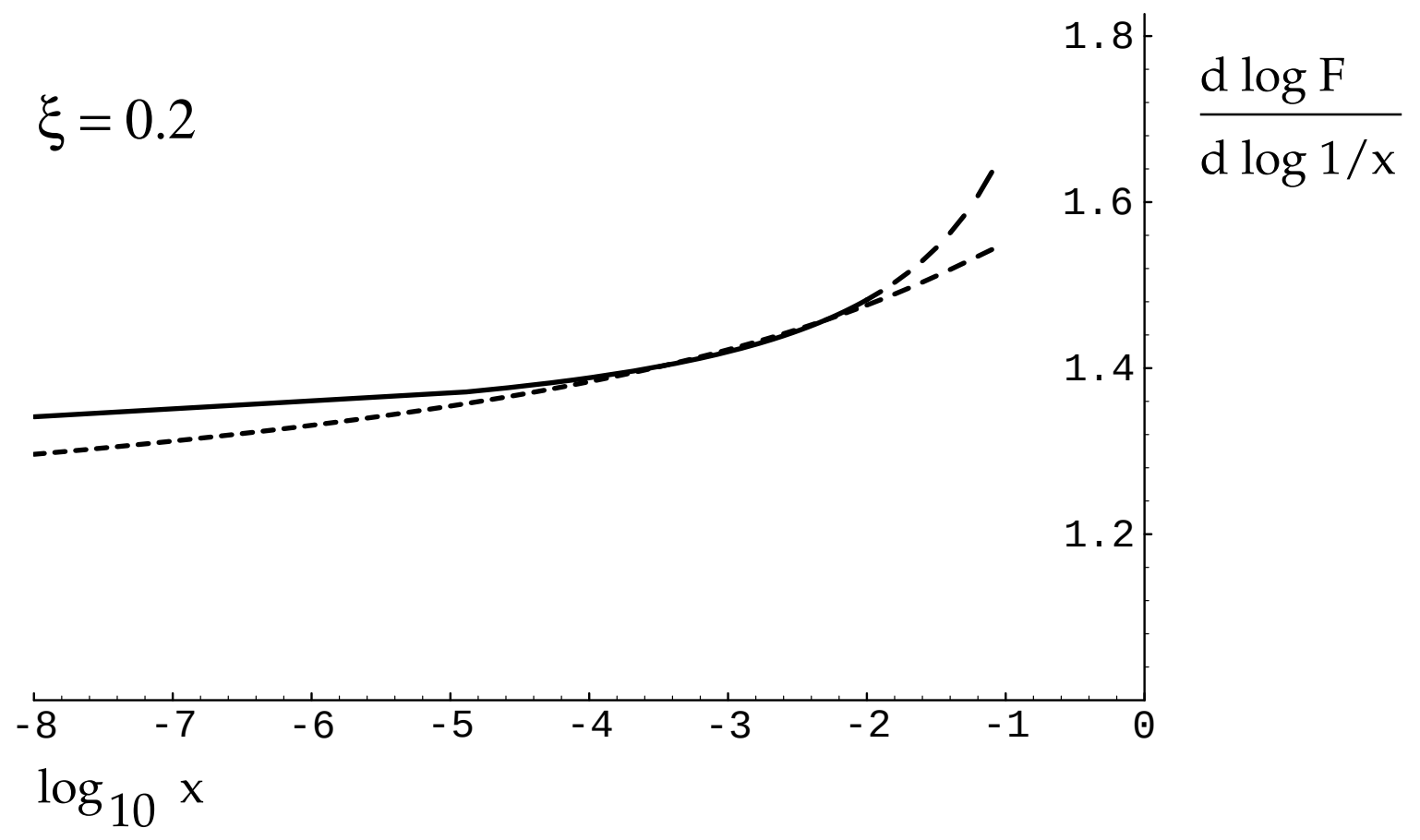
ξ = 0.2
1.8
1.6
1.4
1.2
d log F / d log 1/x
-8 -7 -6 -5 -4 -3 -2 -1 0
log10 x

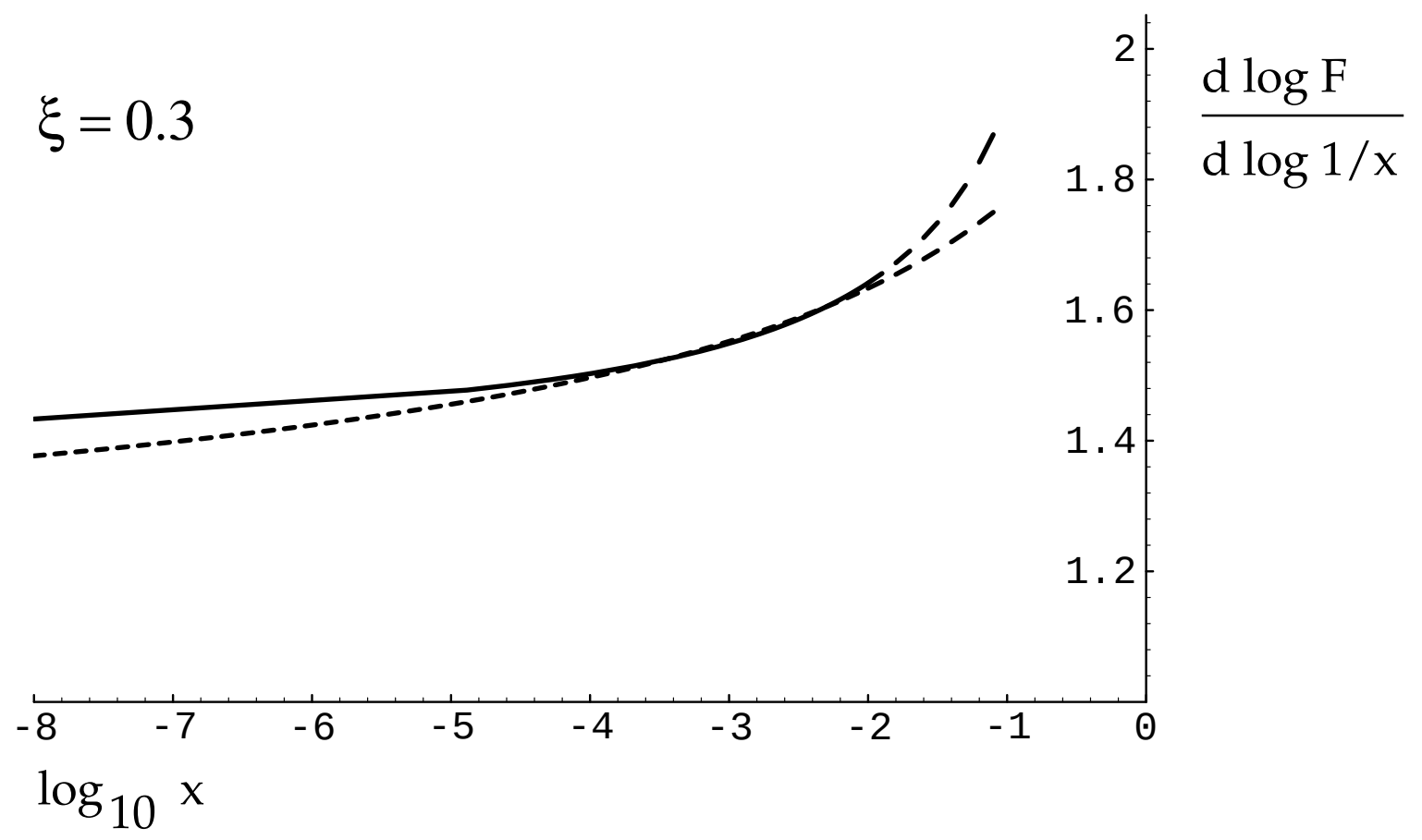
ξ = 0.3
2
1.8
1.6
1.4
1.2
d log F / d log 1/x
-8 -7 -6 -5 -4 -3 -2 -1 0
log10 x

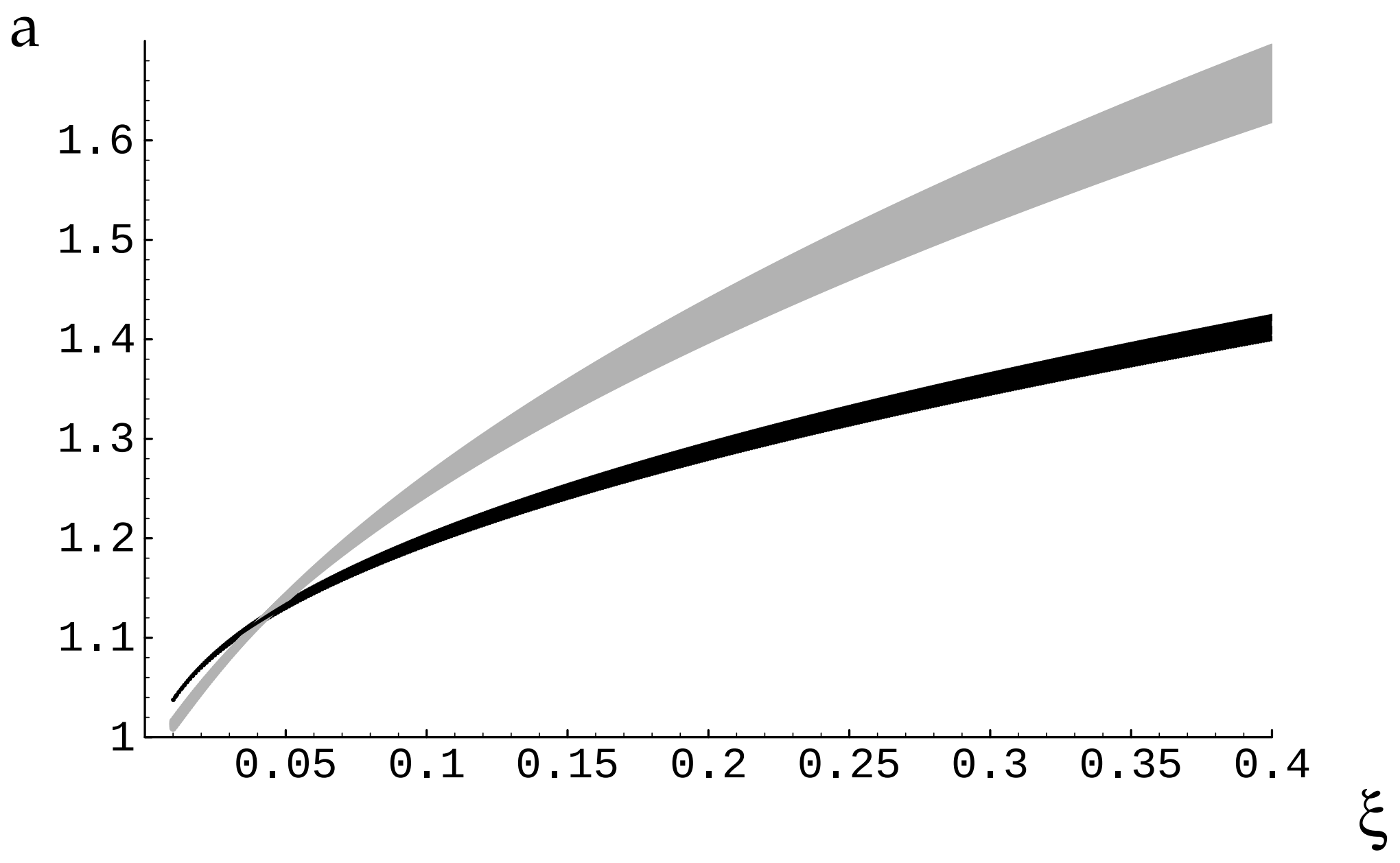
a
1.6
1.5
1.4
1.3
1.2
1.1
1
0.05 0.1 0.15 0.2 0.25 0.3 0.35 0.4
ξ

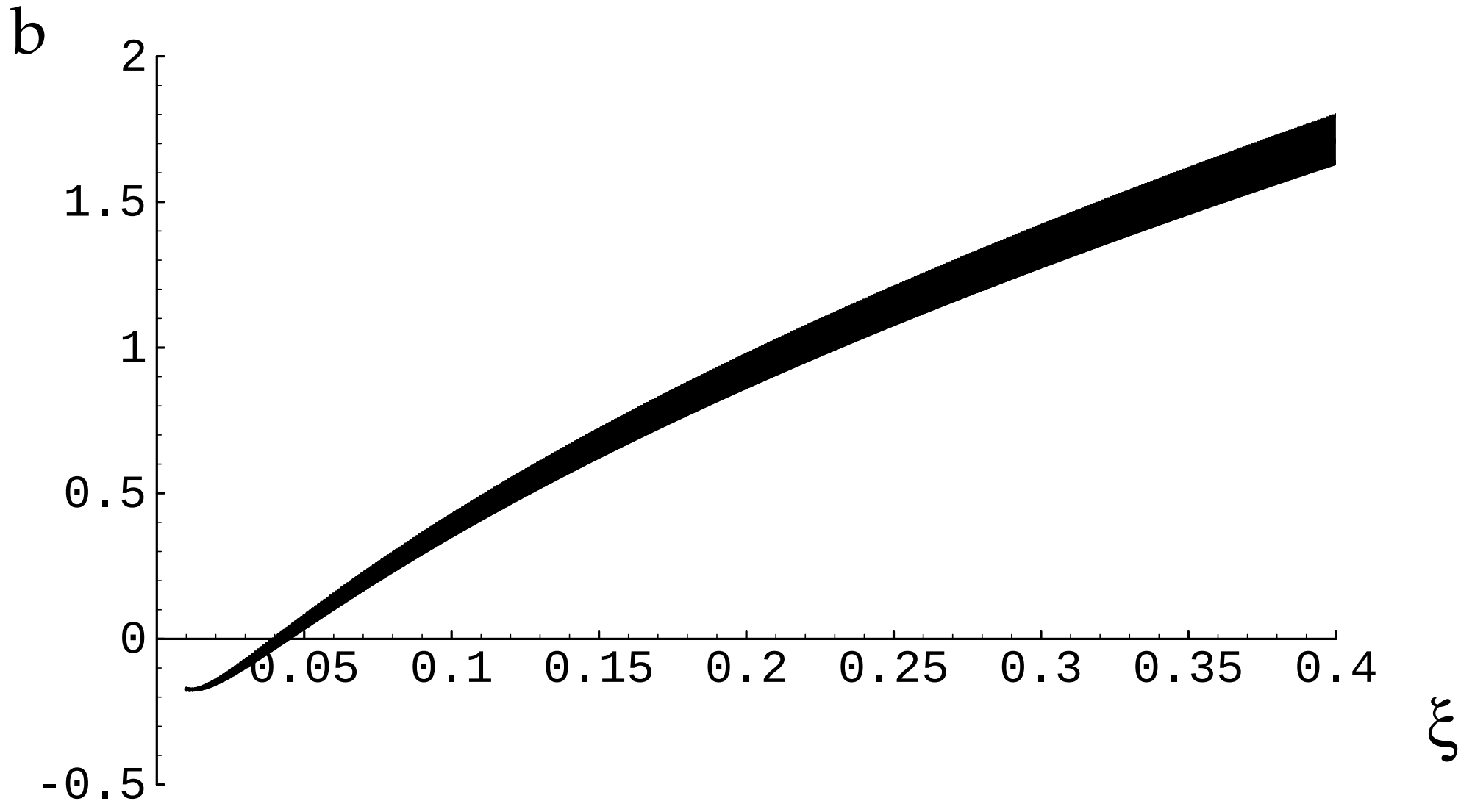
b
2
1.5
1
0.5
0
-0.5
0.05 0.1 0.15 0.2 0.25 0.3 0.35 0.4
ξ

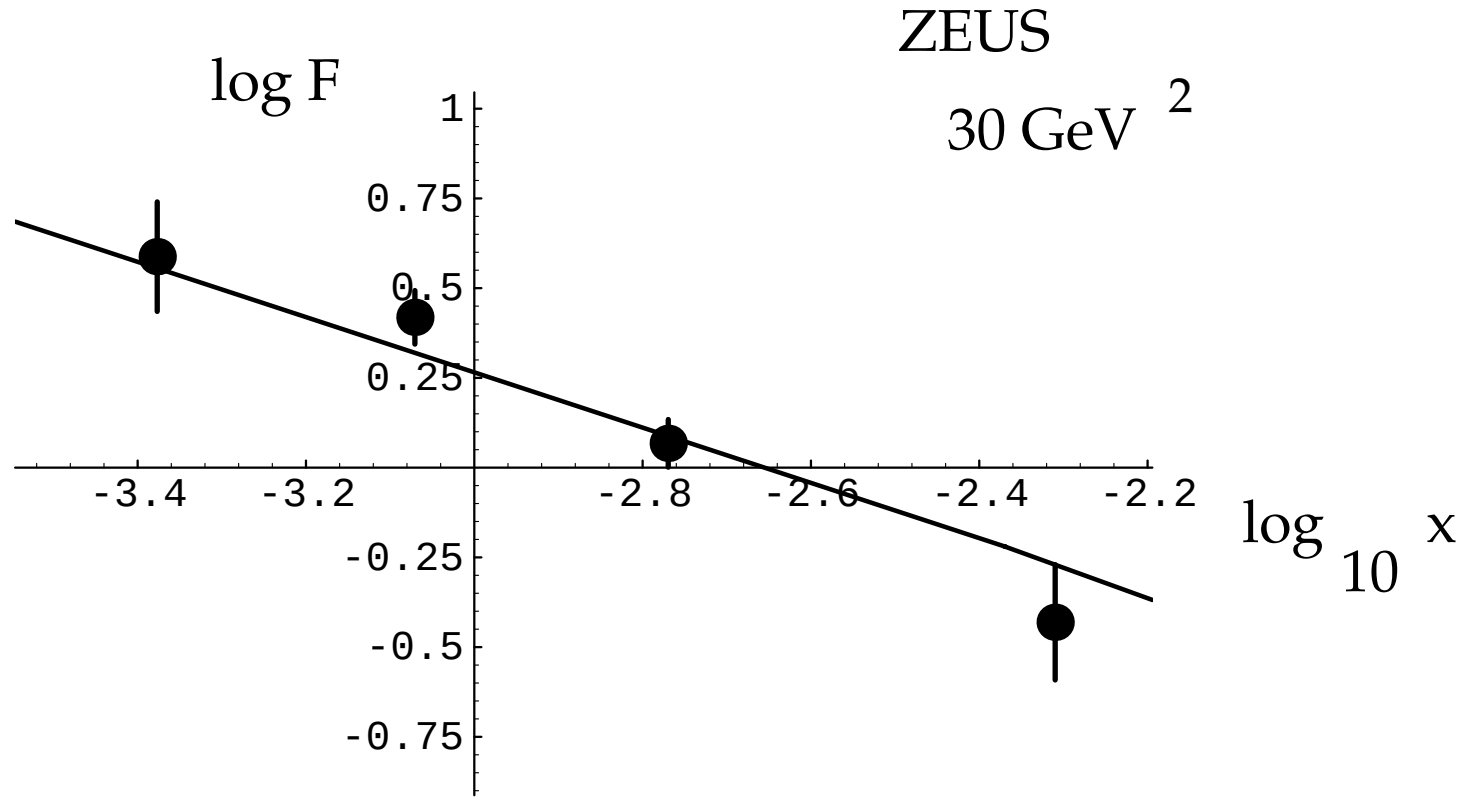

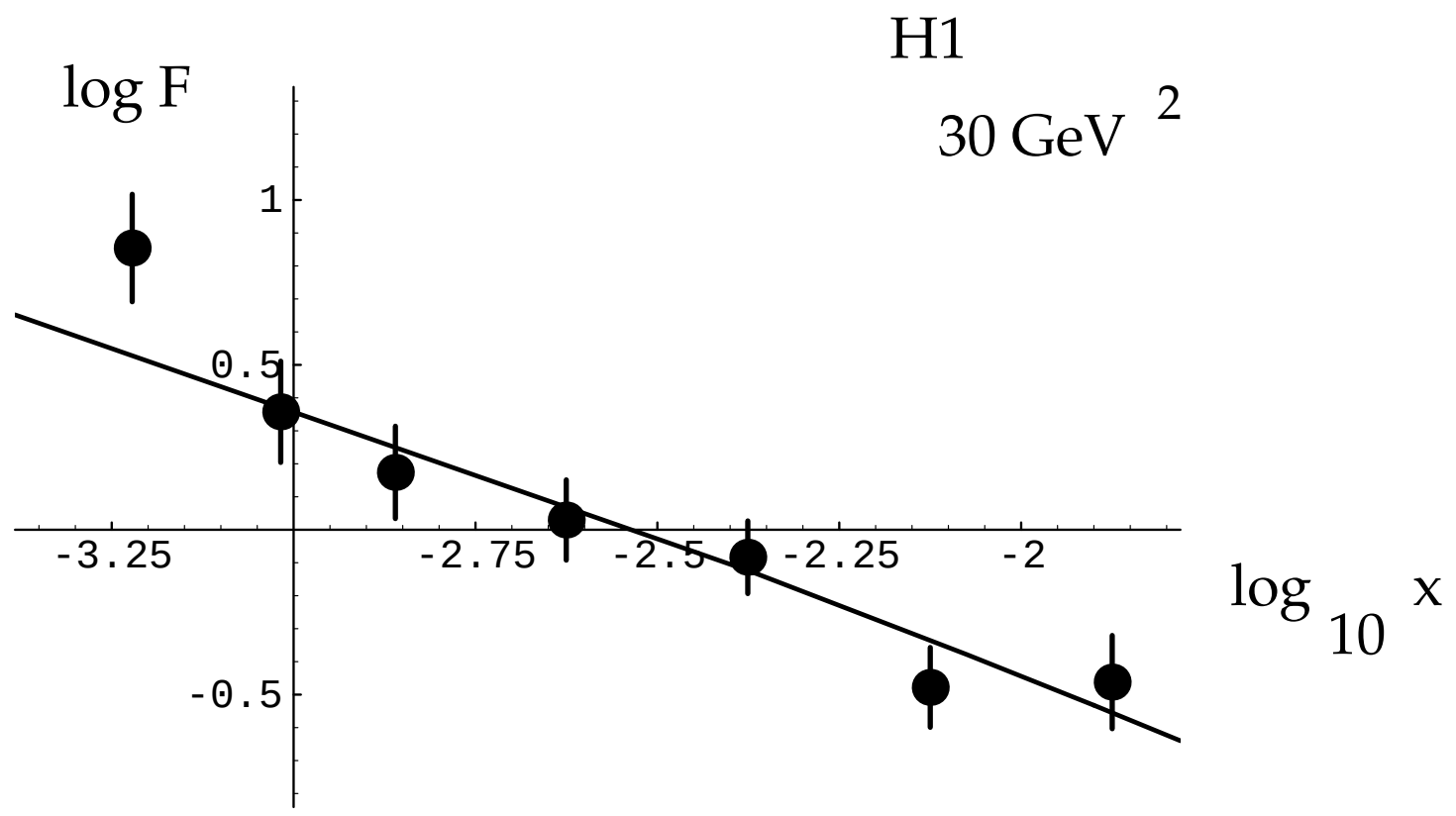

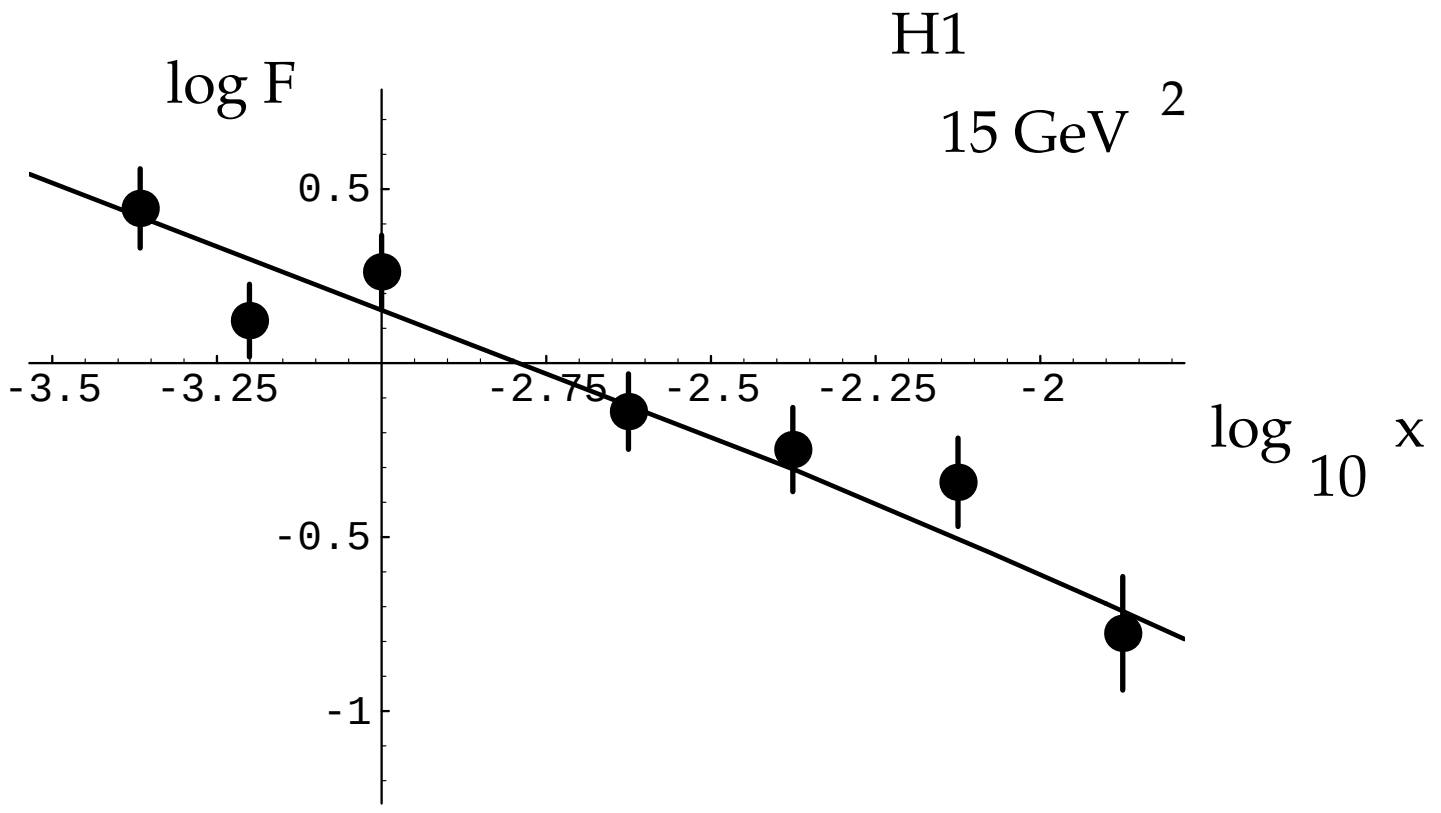